\documentclass[12pt]{iopart}
\input epsf
\newcommand{\be}{\begin{equation}}
\newcommand{\ee}{\end{equation}}
\newcommand{\bea}{\begin{eqnarray}}
\newcommand{\eea}{\end{eqnarray}}
\begin{document}
\title{Circuits in random graphs: from local trees to global loops}

\author{Enzo Marinari$\dag$, R\'emi Monasson${\ddag}$}
\address{\dag Dipartimento di Fisica, SMC-INFM and INFN,
Universit\`a di Roma "La Sapienza",
P. A. Moro 2, 00185 Roma, Italy}
\address{\ddag CNRS-Laboratoire de Physique Th\'{e}orique de l'ENS, 24 rue
Lhomond, 75005 Paris, France}

\begin{abstract}
We compute the number of circuits and of loops with multiple crossings
in random regular graphs. We discuss the importance of this issue for
the validity of the cavity approach. On the one side we obtain
analytic results for the infinite volume limit in agreement with
existing exact results. On the other side we implement a counting
algorithm, enumerate circuits at finite $N$ and draw some general
conclusions about the finite $N$ behavior of the circuits.
\end{abstract}

\section{Introduction}

\begin{figure}
\begin{center}
\mbox{\epsfxsize 2.5in \epsfbox{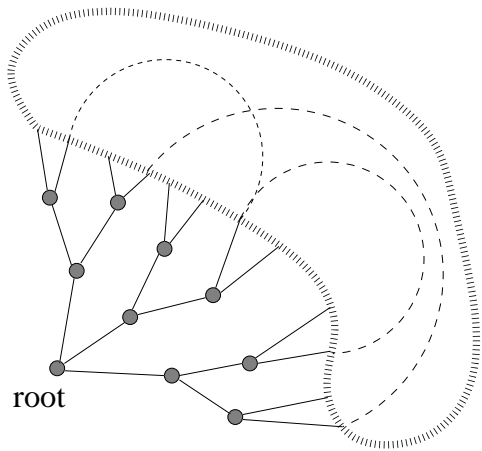}}
\end{center}
\caption{Structure of a random regular graph with degree $c=3$. In the
vicinity of a vertex (the {\em root}) the graph looks like a
(regular) tree attached to the bulk. For any integer $D$, vertices at
distance $\le D$ from the root are almost surely distinct when the
size $N$ of the graph goes to infinity. No short loop, but many long
(dashed) ones pass through the root.}
\label{figtree}
\end{figure}

The study of random graphs, initiated more than four decades ago, has
been since an issue of major interest in probability theory and in
statistical mechanics.  Examples of random graphs abound \cite{boo},
from the original Erd\"os-Renyi model \cite{rg2}, where edges are
chosen independently of each other between pairs of a set of $N$
vertices (with a fixed probability of $O(1/N)$), to the scale-free
graphs with power law degree distribution \cite{bar}, only introduced
in recent times.  An interesting and useful distribution is the one
that generates random $c$-regular graphs \cite{worm1}. These are
uniformly drawn from the set of all graphs over $N$ vertices, each
restricted to have degree $c$. Random regular graphs can be easily
generated when $N$ is large (and $c$ finite) \cite{worm2}, and an
instance of $3$-regular graph is symbolized in figure~\ref{figtree}.
Around a randomly picked up vertex called {\em root} the graph looks
like a regular tree. The probability that there exists a {\em circuit}
(a self-avoiding closed path) of length $L$ going through the root
vanishes when $N$ is sent to infinity and $L$ is kept
finite\footnote{More precisely, this probability asymptotically
departs from zero when $\log N = O(L)$.  Finite-length loops may be
present in the graph, but not in extensive number.}  \cite{worm1}.

The purpose of this article is to reach some quantitative understanding
of the presence of long circuits in random graphs. Our motivation is
at least two-fold.

First, improving our knowledge on circuits in random graphs would 
certainly have positive fall-out on the understanding of equilibrium
properties of models of interacting variables living on
these graphs. For instance, frustration in spin systems emerge from
the presence of circuits with odd length. The number of these circuits
is therefore directly related to the amount of glassiness present in the
system at low temperature. From the dynamical point of view too,
the properties of models of interacting or communicating agents
e.g. routers, ... strongly depend on the topological
properties of the underlying graph, and particularly upon feedbacks 
resulting from the presence of circuits. These ``practical''
incentives, together with purely academic motivations have led to
recent studies on the distributions of unicyclic components \cite{ben}
and cycles \cite{roz} in random graphs.

Secondly, the absence of short circuits in random graphs or,
equivalently, the observation that random graphs essentially exhibit a
local tree-like structure is crucial to the analytical treatment of
spin models on random graphs with some common techniques, e.g. the
replica or cavity methods. These models are characterized by
(realistic) strong interactions between spins, in contrast to usual
mean-field systems, e.g. Curie-Weiss or Sherrington--Kirkpatrick
models. Yet, the absence of short-ranged loops in random graphs make
them, around any vertex (root), locally similar to a tree surrounded
by a remote bulk. Because of that one conjectures that the free-energy
of a spin model on a random graph is equal to the one on the tree with
self-consistent boundary conditions requiring that appropriate
properties (for example the magnetization for the Ising model) are
identical on the leaves and at the root vertex \cite{ded,mott,kan}.
One of the goals of this note is to point out that this procedure {\em
\`a la} Bethe amounts to make assumptions on the large-scale structure
of random graphs, {\em i.e.} on the distribution of long
loops\footnote{This point has certainly already come to mind to
researchers in the field but it is, at the best of our knowledge,
never explicitly mentioned in the literature.}.

We will first discuss {\em loops}, and compute their number in the
thermodynamic limit by analyzing the partition function of the Ising
model.  Next we will compute the number of circuits by discussing the
$n\longrightarrow 0$ limit of the $O(n)$ ferromagnetic model. For
gathering information about the finite $N$ behavior of the number of
loops (or of the loop entropy) we will use an effective counting
algorithm. We will end this note with a discussion of our findings and
of the (many) open issues.

\section{Infinite size limit: mean-field approach}

In this section, we enumerate long loops in random graphs from the
study of associated statistical mechanics models.  We will give our
{\em ad hoc} definition of a {\em loop} in the following section.

\subsection{Loops with multiple crossings}

Let us consider for example Ising spins $S_i = \pm 1$ on vertices
$i=1,\ldots , N$ of the graph of figure~\ref{figtree}, and
ferromagnetic couplings $J_{ij}=1$ on edges $(i,j)$. To calculate the
partition function $Z(\beta)$ of the Ising model at inverse
temperature $\beta$ on the graph of figure~\ref{figtree}, we consider
the leaves of the uncovered local tree {\em i.e.} vertices at distance
$D$ from the root (in figure~\ref{figtree} the maximum drawn distance
is $D=2$) \cite{levin,ded,mott,bac,kan,lef}.  At sufficiently large
$\beta$, a spontaneous, say, positive magnetization $m$ is expected to
be present in the bulk. As a result, the spins attached to the leaves
will feel an external field $H>0$.  Integrating these spins out will
in turn produce an external field $H'$ acting on spins at distance
$D-1$ from the root, with \cite{levin}
\begin{equation}
H' = \frac{(c-1)}{\beta}\,\tanh ^{-1}[ \tanh(\beta) \tanh (\beta H) ]\ .
\end{equation} 
After repeated iterations of this procedure, the field at the root
reaches a stationary value, $H^*$, from which we can derive the
magnetization $m = \tanh [\beta c H^*]$ and the free-energy density
\begin{equation}
f(\beta) = -\frac{c}{2\beta} \ln  2 \big( 
e^{-\beta} + e^{\beta} \, \cosh 2\beta H^*\big) + \frac {c-1}{\beta} 
\ln  2 \cosh \left(\frac{ \beta c  H^* }{c-1} \right)\ .
\end{equation} 
The critical inverse temperature is the smallest value of $\beta$
for which $H^*$ is non zero {\em i.e.} $\beta _c=
\tanh^{-1}(\frac{1}{c-1})$.
 
As is well known, the high temperature expansion of the partition
function $Z$ of the Ising model can be written as a sum over closed
paths going through each vertex an even number of times and each edge
at most once \cite{parisi}. Each path is given a weight $(\tanh \beta
) ^L$ depending upon its length $L$.  In the following we will refer
to such paths as {\em loops}. Gathering all loops with the same weight
{\em i.e.} same length, we obtain
\begin{equation}
Z(\beta ) = 2^N (\cosh \beta)^{\frac{cN}{2}}
\sum _{L} M(L) \; (\tanh \beta )^L\ ,
\end{equation} 
where $M(L)$ is the number of loops of length $L$ that can be drawn
on the graph. It is a sensible assumption that this multiplicity
exhibits an exponential growth with the graph size,
\begin{equation}
M(L=\ell\, N) = \exp [ N \; \sigma (\ell ) +o(N) ]\ ,
\end{equation}
where $\ell$ is the intensive length of the loops, and $\sigma$ is the
entropy of loops having length $\ell$. We stress here that $\ell
=O(1)$ corresponds to $L=O(N)$. Knowledge of the entropy $\sigma$ will
provide us with information about large-scale loops {\em i.e.} with
lengths of the order of $N$. Insertion of the above scaling hypothesis
for $M(L)$ in the partition function $Z$, and use of the Laplace
method yield
\begin{equation}
-\beta f(\beta) = - \ln 2 - \frac c2 \ln \cosh \beta +
\max _{\ell} \big[ \sigma (\ell ) + \ell \; \ln \tanh \beta
\, \big]
 \label{g-sig}
\end{equation}
in the infinite $N$ limit\footnote{Here we are assuming that not only
the free energy density $f$ is self-averaging (as is known and true)
but that also $\sigma$ enjoys the same property.}.  In other words,
the free-energy density $f(\beta)$ is essentially the Legendre
transform of the loop entropy $\sigma (\ell)$.

\begin{figure}
\begin{center}
\hskip .4in A\hskip 2.6in B \hskip 2in\\ \vskip .2 in
\mbox{\epsfxsize 2.5in \epsfbox{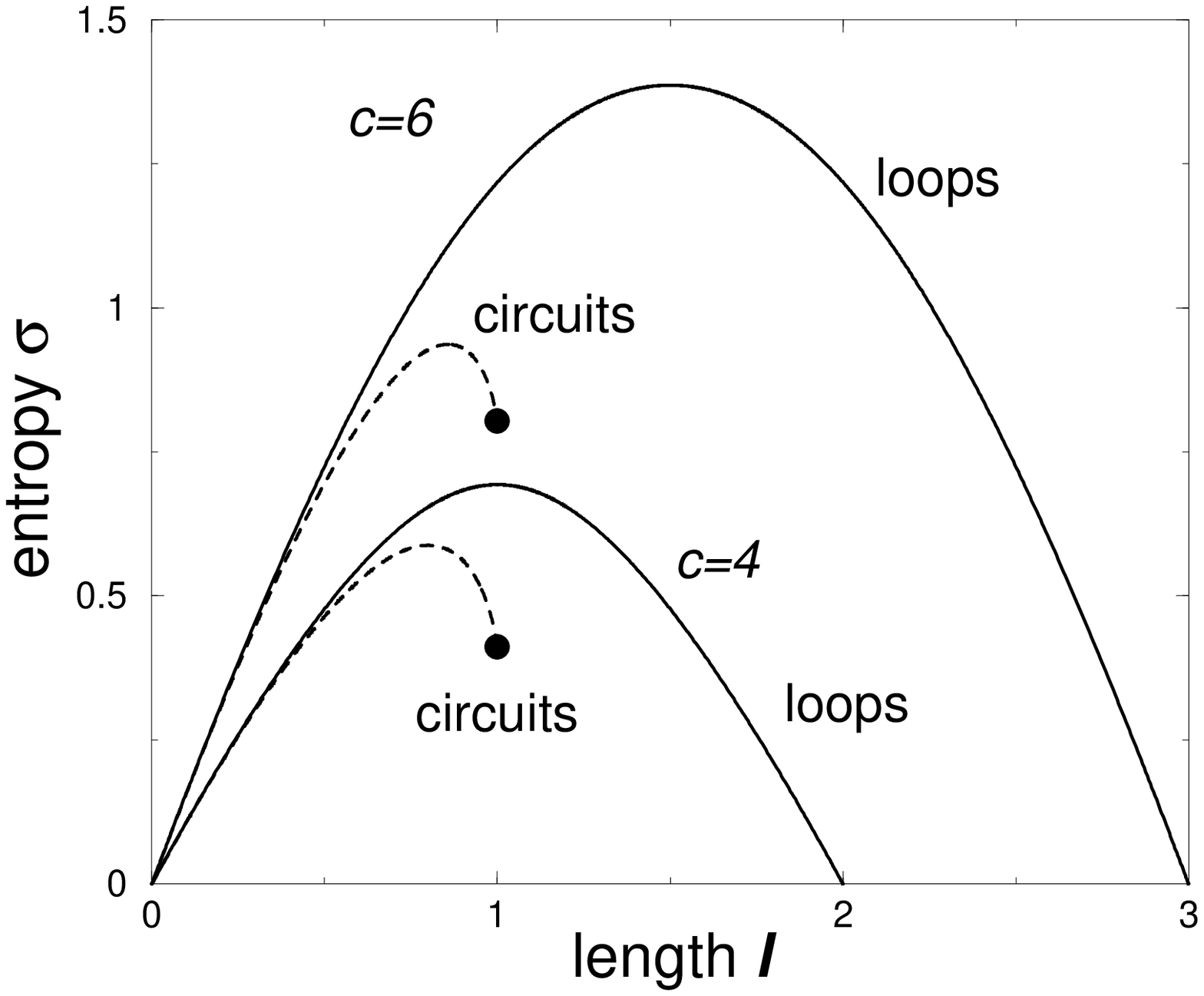}}
\hskip .2in
\mbox{\epsfxsize 2.5in \epsfbox{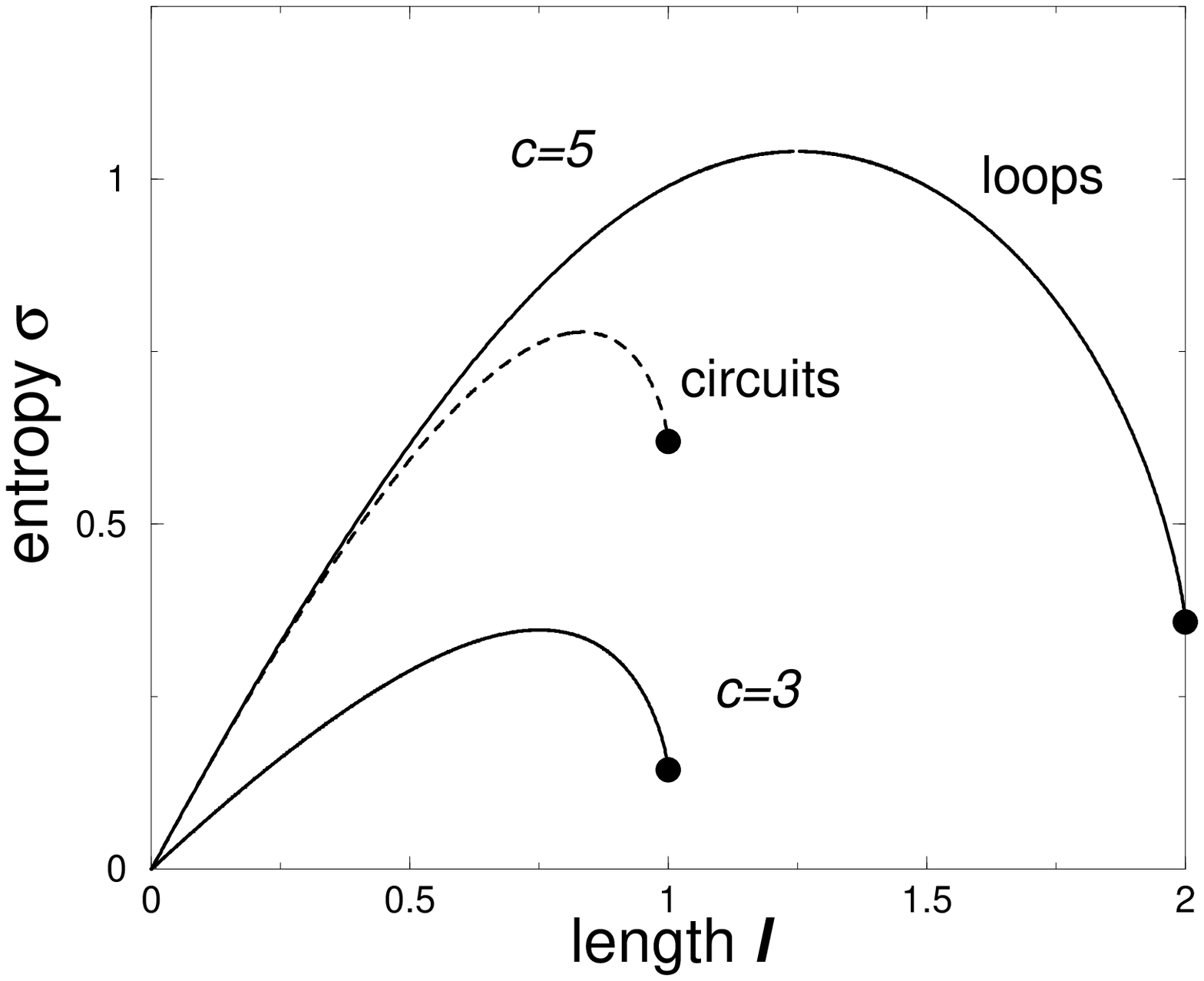}}
\end{center}
\caption{Entropy $\sigma$ as a function of the length $\ell$ of
loops (full curves) and circuits (dashed curves) 
for even ({\bf A}) and odd ({\bf B}) degrees $c$ (for $c=3$, loops
and circuits coincide). The slope at origin is $\ln (c-1)$ for both
cases, the smallest loops being essentially circuits. Black dots,
with coordinates $\ell _{+},\sigma _{+}$ identify 
the longest loops and circuits, see text.} 
\label{figpaire}
\end{figure}

We show in figure~\ref{figpaire} our results for the loop entropy. The
entropy departs from $\ell=\sigma=0$ with a slope equal to $-\ln(\tanh
\beta_c )= \ln (c-1)$ from (\ref{g-sig})\footnote{Note that the finite
slope at origin comes from the fact that the objects counted here
(loops or circuits) are non local; one cannot gain entropy by
rearranging edges independently~\cite{pparisi}.}.  Its maximum is
reached in $\ell_M= \frac{c}4,\sigma _M=(\frac{c}2-1) \ln 2$, that is,
for loops going through half of the edges.  The left part of the
entropy curve ($\ell \le \ell _M$) is parametrized by inverse
temperatures $\beta$ going from $\beta _c$ ($\ell=0$ -- the para/ferro
transition takes place when extensive loops start contributing to
partition function $Z$) to $\infty$ (top of the curve -- at zero
temperature, $Z$ is dominated by the most numerous paths).  The right
part of the entropy curve ($\ell \ge \ell _M$) is obtained when $\tanh
\beta >1$, that is, for inverse temperatures with an imaginary part
equal to $\frac{\pi}2$ allowing to transform the hyperbolic tangent of
$\beta$ into the cotangent of its real part. This is done by
considering a real $\beta > \beta _c$, the stationary field $H^*$ for
this temperature, and the transformation $\beta \to \beta + i
\frac{\pi}2$, $(\beta H)^*\to (\beta H)^* +i (c-1) \frac{\pi}2$.

{\em The case of even degree.} For even $c$ $\sigma$ is left unchanged
under the transformation $\ell \to \frac{c}2-\ell$.  The right part of
the entropy curve is thus the mirror symmetric of the left part with
respect to $\ell= \ell _M$ (figure~\ref{figpaire}A).  This results
from a duality between long and short extensive loops.  Let $\Gamma
^F$ be the full loop (all edges occupied, which is allowed for even
$c$). Then, for any loop $\Gamma$ with length $\ell< \frac{c}4$,
$\Gamma ^F \setminus \Gamma$ (that is $\Gamma ^F$ without the edges
belonging to $\Gamma$) is a loop with length $\frac{c}2-\ell$. Notice
that the largest loop is $\Gamma^F$, with length $\frac{c}{2}$.

{\em The case of odd degree.} Here duality does not hold and there is
no simple transformation rule for $\ell$ and $\sigma$.  An explicit
calculation gives the entropy curve shown in
figure~\ref{figpaire}B. The maximal length $\ell_+ =\frac{(c-1)}{2}$
is reached with an infinite slope, and corresponds to a finite entropy
$\sigma _+ =\frac12 ( (c-1) \, \ln (c-1) - (c-2) \ln c)$.  For odd
degrees $c$, indeed, loops cannot occupy all edges since the number of
incoming edges onto each vertex is even.  The longest loops have one
free edge per vertex: these act as defects, the positions of which can
be chosen with some freedom, giving rise to a finite entropy.  The
larger is $c$, the smaller is the ratio $\frac{\sigma_+}{\sigma_M}$:
intuitively, the frustration coming from the parity of $c$ is less and
less important as $c$ increases\footnote{Numerical investigations of
the one-step replica symmetry broken Ansatz for the Ising model
with $\tanh \beta >1$ indicate that the replica symmetric solution is
correct in this range of (complex-valued) temperature \cite{pparisi}.}.

\subsection{Loops without crossings: circuits}

The iteration procedure can be applied to calculate the free-energy of
other short-range models and then derive the entropies corresponding
to large-scale diagrams generated through high temperature expansions.
For instance the ferromagnetic $O(n)$ model with $n\to 0$ gives
information on circuits {\em i.e} loops with vertices having degree 2
at most \cite{deg}. A spin $\vec S$ is submitted to two fields
$H_1,H_2$ conjugated to the magnetization $\frac{\vec S \cdot \vec
1}{\sqrt n}$ and its squared value\footnote{Higher moments are not
relevant in the $n\to 0$ limit.}. The iteration equations for these
fields read
\begin{eqnarray}
H_1'&=& \frac{(c-1) \beta\, H_1}{1+\,H_2}\ , \\
\nonumber
H_2'&=& \frac{(c-1)(c-2)\beta^2 H_1^2}{2(1+\,H_2)^2}\ .
\end{eqnarray}
Inserting the fixed point values for the fields in the expression of
the free-energy per spin component,
\begin{equation}
f = \frac {1}{\beta} \ln \left[ (1+H_2)^{c} \left(
1 + \frac{c(c-1) \beta^2 H_1^2}{2 (1+H_2) ^{2}}\right) \right] 
- \frac{c}{2\beta} \ln \big[ (1+H_2)^2 + \beta H_1^2 \big] \ ,
\end{equation}
yields (for an alternative derivation, see \cite{poly}),
\begin{equation}
f = -\frac {c-2}{2\beta} \ln \left[ \frac{c 
(c-1)\beta -2}{c-2} \right] + \frac{c}{2\beta} \ln \big[
(c-1)\beta\big] \ .
\end{equation}
The entropy of circuits,
\begin{equation}
\label{mc}
\sigma (\ell ) = - (1-\ell) \ln (1-\ell ) + \left( \frac c2 - \ell \right)
\, \ln \left(1 - \frac{2\ell}c\right) + \ell \,\ln (c-1) \ ,
\end{equation}
is obtained through $-\beta f(\beta) = \max _\ell [\sigma (\ell) + \ell \ln
\beta]$.  Results are shown in figure~\ref{figpaire}.  The maximal
entropy, $\sigma _M =\frac12 (c \ln (c-1)-(c-2) \ln (c+1))$, is
reached in $\ell _M = \frac{c}{(c+1)}$. The rightmost point, with
coordinates $\ell_+=1,\sigma _+ = \ln (c-1) + \frac12 (c-2) \ln
(1-2/c)$ corresponds to Hamiltonian cycles. Expression (\ref{mc})
coincides with the output of rigorous calculations \cite{rg2,ga},
which shows the exactness of the replica symmetric hypothesis for the
$O(n\to 0)$ model \cite{poly}. Note the similarity of the entropy
curves for loops on odd $c$-regular graphs and circuits
(figure~\ref{figpaire}).  In the two cases the slopes at minimal and
maximal lengths are respectively finite and infinite, see the
discussion above.

Let $c\ge 3$ and $M(L)$ be in all the rest of this note
the number of circuits of length $L$ in a random, $c$-regular graph.
For finite $L$, $M(L)$ is asymptotically Poisson-distributed when
$N\to \infty$ \cite{worm1},  
\begin{equation}
\protect\label{eq-poisson}
{\rm Proba}[ M(L) = M]  = \frac1{M!}
\left[  \frac{\left(c-1\right)^L}{2L} \right]^{M}
\exp\left\{-  \frac{\left(c-1\right)^L}{2L} \right\}\ .
\end{equation}
The above identity holds indeed for circuit-length $L \ll \log N$.
The expected number of circuits of intensive length $\ell= L/N$ is
therefore, for $\ell<\frac{\log N}{N}$,
\begin{equation}
\label{form}
\langle M(\ell) \rangle =  \frac{\left(c-1\right)^L}{2L}= \exp
\left\{ N \sigma(\ell) -\log(N)+\tilde{\sigma}(\ell)
\right\}\ ,
\end{equation}
with $\sigma(\ell)=\ell\ \log(c-1)$ and
$\tilde{\sigma}(\ell)=\log(2\ell)$. Note that this value of $\sigma
(\ell)$ coincides with expression (\ref{mc}) at the first order in $\ell$.

\section{Finite size corrections to circuit number: numerical results}

To study what happens at finite $N$ we have implemented a fast
algorithm for finding all circuits in a given graph~\cite{algo}. It is
important that we explicitly {\em find}, not only count, all the
circuits so that by our method we can in principle give all
interesting characterizations.

In our enumeration we first generate a random graph and then count the
circuits. We average this outcome over a number of samples. We
describe the algorithm for enumeration in next section.

To generate a fixed connectivity random graph we start by assuming
that each site has $c$ connections that connect it to $c$ different
sites: we allow no self-connections and no double edges. At the
beginning, all connections are free. Two of the connections are
extracted and matched together. We continue filling them up (we use a
table that we resize at each step to keep the process effective) till
all connections are set or we are stuck because there are two left
connections that both belong to the same site or that belong to two
sites that are already connected. To be sure not to introduce any
systematic bias in this case we just discard the full graph and
restart the procedure from scratch.

\subsection{An algorithm for circuit enumeration}

For enumerating circuits we have implemented and used an algorithm
introduced by Johnson~\cite{algo}. The algorithm finds all elementary
circuits of a graph. The computer time needed for this task is bounded
by $O((N+E)(M+1))$, where $N$ is the number of vertices of the graph,
$E$ the number of edges and $M$ the total number of circuits in the
graph: indeed one proves that the time used between the output of two
consecutive circuits is bound by $O(N+E)$ (and that this is true also
for time elapse before the output of the first circuit and after the
output of the last one).  The memory needed by the algorithm is bounded
by $O(N+E)$.

Let us just sketch the crucial steps of the procedure. One first orders
the vertices in some lexicographic sequence, and labels them with
integers ranging from $0$ to $N-1$. The search is started from a {\em
root} vertex $r$, in the subgraph induced by $r$ and by vertices after
$r$: one of the crucial performance issues is that all vertices that
become roots are the smallest vertex in at least one elementary
circuit. The input to the procedure is the adjacency structure $A(v)$
for each vertex $v$, that represents the graph: it contains $u$ if and
only if $(v,u)\in {\cal E}$, where $\cal E$ is the set of edges of the
graph.

We {\em block} a vertex $v$ when it is added to a path
beginning in $r$ . We build elementary paths starting from $r$. The
vertices of the current trial paths are loaded on a stack. A procedure
adds the vertex to the path, if appropriate, and appends the vertex to
the stack: the vertex is deleted from the stack when returning from
this procedure. The ingenious part of the algorithm is in keeping a
vertex {\em blocked} as long as possible, to minimize the execution
time. This has to be done while keeping the procedure correct: the
basic rule that has to be satisfied to guarantee that all circuits are
found (only once) is that if it exists a path from the vertex $v$ to
$r$ that does not intersect the path loaded on the stack, then $v$ has
to be free (i.e. it cannot be in a {\em blocked} state). One uses a
list to keep the information needed to satisfy this constraint while
staying time effective.

Some details about performances are as follows: on an Intel Xeon $2.8$
$GHz$ processor our implementation takes of the order of $0.07$
seconds for finding all circuits of a $N=30$ graph (they are
$O(50000)$), $2.4s$ for $N=40$ ($O(1.5\ 10^6)$ circuits) and $80s$ for
$N=50$ (with $O(4\ 10^7)$ circuits).

In order to test the results of our code we have also implemented a
``quick and dirty'' backtracking procedure to count Hamiltonian
circuits. Since our procedure crucially depends on the quality of the
random number generator we have also checked that different (high
quality) random number generators lead to statistically compatible
answers. 

\subsection{Results and interpretation}

Thanks to our algorithm and implementation we have been able to
enumerate of the order of $10^{14}$ circuits (a large number). 

For small $c$ values we can study larger graphs (we have analyzed
graphs with up to $64$ vertices in the $c=3$ case and up to $22$
vertices for $c=6$, and averaged our results over samples ranging from
$1000$ to $10000$ random graphs). Typically we find for example of the
order of $300$ million circuits on a $N=56$, $c=3$ graph, one billion
circuits on a $N=26$, $c=5$ graph and $1.5$ billion circuits on a
$N=22$, $c=6$ graph. For each value of $N$, we average over of the
order of $10000$ samples for all the $c=3$ enumerations, and $1000$
graphs for $c>3$.

\begin{figure}
\begin{center}
\mbox{\epsfxsize 4.0in \epsfbox{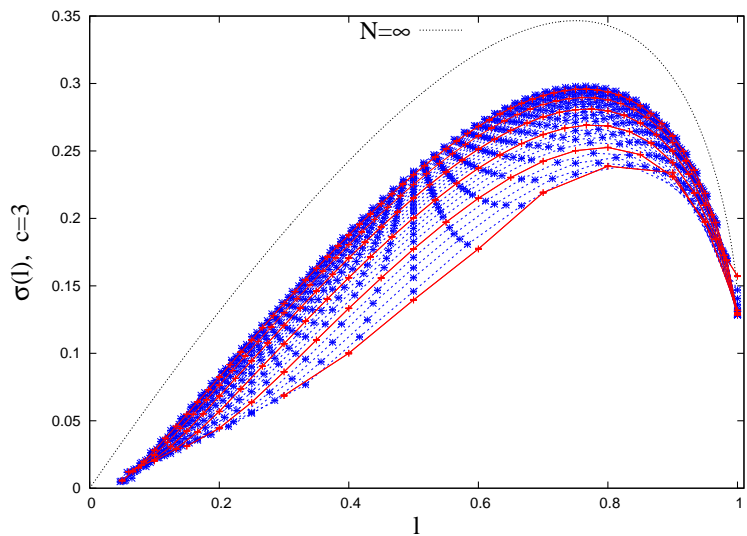}}
\end{center}
\caption{Circuit entropy $\sigma _N(\ell)$ as a function of $\ell$ for 
$c=3$, and for graph sizes
ranging from $N=10$ to $64$ (from bottom to top). The full curve
corresponds to the analytical calculation of equation (\ref{mc}). Data
for sizes multiple of $10$ are made visible by using a different
drawing style.}
\label{figallc3}
\end{figure}

In figure~\ref{figallc3} we plot the theoretical results obtained in
the (replica symmetric) calculation of Section 2.2 for $c=3$, together
with results obtained by finite enumeration on finite lattices with
$N$ vertices.  Numerical data from finite graphs are very slowly
approaching the theoretical, infinite volume data. We try in the
following to analyze the quality of this asymptotic agreement of the
two sets of data.

At first we know that $\log \langle M(\ell) \rangle \sim \ell
\,\log(c-1)$ as $\ell\to 0$ where $\langle M(\ell)\rangle$ is the
average number of circuits of length $L=\ell\cdot N$.  With our
numerical data we cannot work really close to $\ell=0$, since for
finite $N$ we have a value $\ell_{min}$ that goes to zero as $N\to
\infty$: finite size effects appear as a flattening of $\log\langle
M(\ell)\rangle$ when $\ell$ becomes very close to the minimum allowed
value (see figure \ref{figallc3}).  One can easily see by eye that the
slope is very similar to the asymptotic slope in the small $\ell$
region where we are relatively safe from finite size effects.  We have
fitted a linear behavior (that is indeed clear in the data) for
example for $\ell$ in the range $(.13,.19)$ for $c=3$.

Using this approach we find for $c$ from 3 to 6 slopes about $20\%$
smaller than the theoretical prediction (on the larger graphs we can
study): for $c=3$ we find $0.54$ versus a theoretical $\log 2 \sim
0.69$; for $c=4$ we find $0.87$ versus $1.10$; for $c=5$ we find
$1.12$ versus $1.39$; for $c=6$ we find $1.31$ versus $1.61$. Finite
size effect can be drastically reduced if we compare directly
different $c$ value (since we are using graphs of a size that is
limited by the same criterion for each $c$ value). For example the
ratio of the slope of $c$ and $c+1$ is $0.62$ for $c=3$ versus a
theoretical $0.63$, $0.78$ versus $0.79$ for $c=4$ and $0.85$ versus
$0.86$ for $c=5$. This really remarkable agreement gives us confidence
that we have a good control over finite size effects.

\begin{figure}
\begin{center}
\mbox{\epsfxsize 4.0in \epsfbox{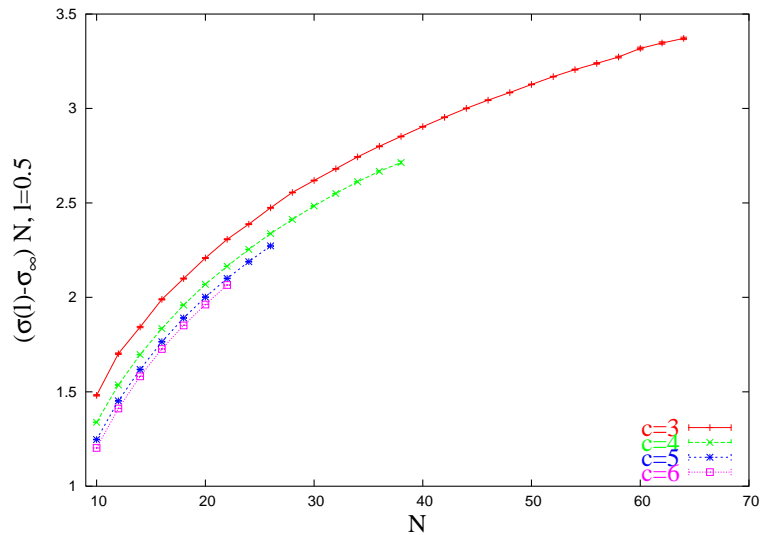}}
\end{center}
\caption{$N$ times the difference between the circuit entropy and its 
asymptotic value, as a function of $N$ for different connectivities $c$. 
Here $\ell=0.5$.}
\label{figallcl05}
\end{figure}

\begin{figure}
\begin{center}
\mbox{\epsfxsize 4.0in \epsfbox{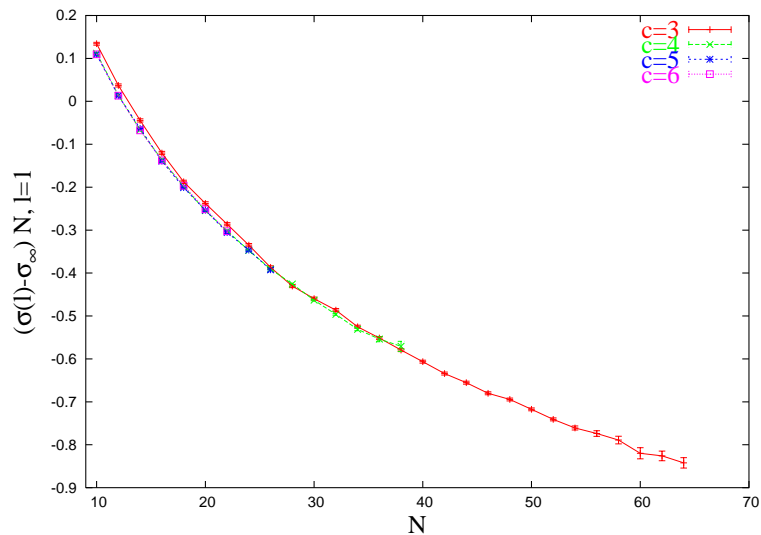}}
\end{center}
\caption{Same as in figure~\ref{figallcl05}  but for Hamiltonian circuits
 {\em i.e.} $\ell=1$.}
\label{figallcl1}
\end{figure}

Equation (\ref{form}) tells us that the difference between  
the measured circuit entropy $\sigma _N (\ell)$ and the asymptotic
value given by formula (\ref{mc}), once multiplied by $N$, behaves as
\begin{equation} 
\label{mc1}
\big(\sigma _N(\ell) - \sigma _\infty (\ell) \big) N = - \log N
+ \tilde \sigma (\ell)
\end{equation} 
for vanishing small values of $\ell$, with $\tilde \sigma (\ell) =
-\log (2\ell)$. It is therefore independent of $c$, with a logarithmic
dependence upon the graph size $N$. To check how equation (\ref{mc1})
applies to finite values of $\ell$, we look first in
figure~\ref{figallcl05} at the number of circuits with $\ell=0.5$, for
different values of $N$ and $c$ (we have harvested our most precise
data at $c=3$).  Data indeed show only a very weak dependence upon
$c$, and this dependence becomes weaker with increasing $c$. Data for
$c=5$ are already indistinguishable from the ones for $c=6$.

In figure~\ref{figallcl1}, we show the same quantity for $\ell=1$ {\em
i.e.} for Hamiltonian circuits, that pass through all vertices of the
random graph. As far as the scaling with $c$ is concerned
figure~\ref{figallcl1} shows that the scaling of Hamiltonian circuits
is excellent already at $c=3$.  We will come back later about the fact
that scaling property of Hamiltonian circuits turn out to be very
different from the ones of all other finite $\ell$, less dense
circuits.

\begin{figure}
\begin{center}
\mbox{\epsfxsize 4.0in \epsfbox{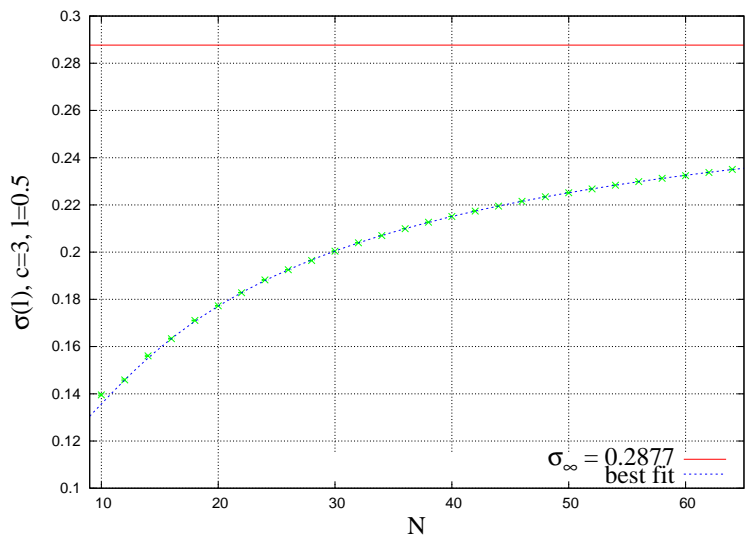}}
\end{center}
\caption{Circuit entropy $\sigma(\ell=0.5)$ versus $N$ for $c=3$ and 
best fit to the form (\ref{ourbestfit}).}
\label{figfit}
\end{figure}

Inspired by the result obtained for small circuits
(\ref{form}),(\ref{mc1}) we look for the following fit for the circuit
entropy for finite values of $\ell$,
\begin{equation}
\sigma_N(l) = \sigma_\infty(l) + c_1 \frac{\log N }{N} + c_2 \frac1N\ . 
\label{ourbestfit}
\end{equation}
In figure~\ref{figfit} we show our results for $c=3$, $\ell=0.5$.
The quality of the best fit to data with sizes
$N\ge 30$ only is excellent, and is very good agreement with all data
with $N \ge12$. The two
parameter fit is clearly superior to power law fits.
We find that with very good accuracy (surely better than one percent) 
$$
c_1=-1\ ,
$$
i.e. that even at finite $\ell$ the relation (\ref{mc1}) gives the correct
leading corrections. This result tells us that
for all $\ell$ values (maybe excluding $\ell=1$, see later) we have
that the average number of circuits of reduced length $\ell$ equals
\begin{equation} 
\label{mc2}
\langle M(\ell \, N) \rangle = (K(\ell) + o(1) \big) \; \frac{e^{N \, \sigma
    _\infty(\ell)}}{N}\ ,
\end{equation}
where $K(\ell)$ is a bounded function of $\ell$.  
For $c_2$, we find values close to
$1$ e.g. $.78$ in the case of $\ell=0.5$. Here precision is not as good
since this is a sub-leading correction. What is clear from our data is
that sub-leading corrections to the circuit entropy are of the order of 
$1/N$ as encoded in formula (\ref{mc2}).

\begin{figure}
\begin{center}
\mbox{\epsfxsize 4.0in \epsfbox{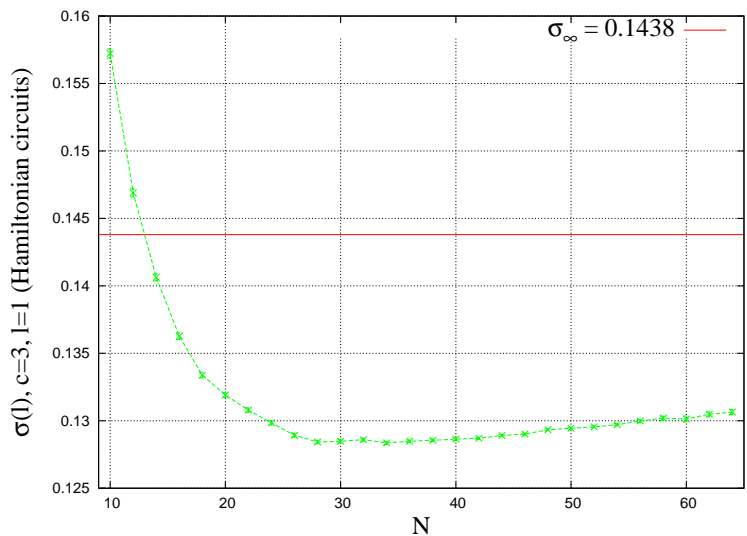}}
\end{center}
\caption{$\sigma(l=1)$ versus $N$ for $c=3$.}
\label{fignofit}
\end{figure}

As we have said above the case of Hamiltonian circuits ($\ell=1$) is
exceptional. Finite size transient effects are very strong; this is
intuitively expected since these circuits fill the graph and are
deeply affected by its finite size. We show in figure~\ref{fignofit}
the analogous of figure~\ref{figfit} but for $\ell=1$. It is clear
that the structure of finite size effects is completely different. On
the contrary we have already explained that we find exactly the same
behavior for all intermediate $\ell$ values: the case $\ell=1$ appears
as isolated.

\section{Conclusions}

To sum up, the absence of small loops in random graphs allows one to
argue that the free-energy of a spin model defined on the graph is
equal to the one on a regular tree with a self-consistent external
field at boundary (leaves). In turn, this free-energy fully determines
the distribution of large-scale loops in the random graph. We stress
that the whole complexity of this cavity procedure is actually hidden
in the assumption of self-consistency for the field(s) \cite{thou}.
Assumptions corresponding to broken symmetry in the replica framework
\cite{mon,mez} may be necessary for models with random
interactions. Anyhow, the salient feature of the above standard
approach is that it predicts global thermodynamical quantities from
local considerations about the graph only.

We have added to our exact computation, valid in the $N
\longrightarrow \infty$ limit, results from exact enumeration at
finite $N$.  Thanks to them we have been able to determine precisely
the behavior of the leading corrections to the thermodynamical
behavior (at least for circuits with $\ell < 1$: we have found
that Hamiltonian circuits have stronger finite size corrections and a
peculiar finite $N$ behavior.

It would be very interesting to characterize properties or quantities
that can adequately be described by a local procedure \cite{aldous}.
The question is however well beyond the scope of the present paper,
which only shows that large loops are (remarkably) among those ones.

\vskip 1cm {\bf Acknowledgments} \vskip .5cm

We are grateful to G. Parisi for insights about the validity of
replica symmetry for the Ising model with complex temperature.
We also thank S. Cocco, A. Montanari for interesting discussions.
R.M. thanks the Les Houches summer school and the SMC-INFM Center in
Rome for the hospitality in August and September 2003 respectively.
E.M. thanks the ``Laboratoire de Physique Th\'eorique'' of ENS in
Paris for a visiting professorship during May-June 2004.

\vskip 1cm {\bf References} \vskip .5cm


\end{document}